\documentclass[%
 reprint,
 amsmath,amssymb,
 aps,
]{revtex4-2}
\usepackage{tikz}
\usepackage{placeins}
\usepackage{quantikz}
\usetikzlibrary{positioning}
\usepackage{babel}
\usepackage{siunitx}
\usetikzlibrary{calc}
\usepackage{mathtools}
\usepackage{amsmath}

\DeclareMathOperator*{\argmin}{arg\,min}
\usepackage{amssymb}
\usepackage{graphicx}
\usepackage{dcolumn}
\usepackage{bm}
\usepackage{hyperref}
\usepackage[mathlines]{lineno}

\usepackage{subcaption}
\usepackage{placeins}
\begin{document}

\preprint{}

\title{Excited-CAFQA: A classical simulation bootstrap for the variational estimation of molecular excited states
}

\author{Bikrant Bhattacharyya}
 \affiliation{California Institute of Technology}
 \email{bbhattac@caltech.edu}
\author{Gokul Subramanian Ravi}%
 \email{gsravi@umich.edu}
\affiliation{%
 University of Michigan
}%

\date{\today}
\begin{abstract}
Variational Quantum Algorithms (VQAs) are iterative algorithms suited to implementation on current-era quantum devices. VQAs employ classical optimization to minimize cost functions evaluated on quantum circuits. However, the extent to which VQAs manage noise is often insufficient for quantum chemistry applications. One method of improving VQAs is through accurate ansatz initialization. The CAFQA (Clifford Ansatz For Quantum Accuracy) protocol runs a discrete search through a classically simulatable subset of the entire state space to find a desirable initialization. Prior work has evaluated CAFQA applied to the Variational Quantum Eigensolver (VQE), a VQA that computes grounds states of a Hamiltonian. Motivated by CAFQA's success, we propose Excited-CAFQA initialization for Variational Quantum Deflation (VQD), a quantum algorithm that extends VQE by allowing the computation of excited states. VQD recursively computes excited states, by constraining the kth state to be orthogonal to the previous $k-1$ computed energy states via a penalty term appended to the standard VQE cost function. Just as with VQE, the VQD cost function can be efficiently computed classically for the states considered in the discrete CAFQA search, allowing for the discrete CAFQA optimizer to find good initial parameters for each energy level computation. Preliminary evaluation shows that Excited-CAFQA achieves $90$ to $99 +\%$ accuracy across a variety of bond lengths and excited states for $\textrm{H}_2$ and $\textrm{HeH}^+$ molecular systems. 
\end{abstract}

\maketitle
\section{Introduction}
\subsection{The Variational Quantum Eigensolver}
The Varational Quantum Eigensolver relies on the \textit{Variational Principle}, which states that for a given Hamiltonian $H$ and an arbitrary quantum state $|\psi\rangle$, \begin{equation}
\begin{aligned}
E_0\leq \langle \psi | H | \psi \rangle 
\\
|\psi_0\rangle = \argmin \langle \psi | H | \psi \rangle.
\end{aligned}
\end{equation}
Where $E_0$ and $|\psi_0\rangle$ are the ground energy and state of $H$ respectively \cite{Tilly_2022}. 
\par
Thus, if $|\psi\rangle$ can be parameterized (with sufficient expressibility) as $|\psi(\theta)\rangle$, then 
\begin{equation}
   \begin{aligned}
    E_0\approx \min_\theta \langle \psi(\theta) | H | \psi(\theta)\rangle
    \\
    |\psi_0\rangle\approx |\psi(\theta^\ast)\rangle \quad \theta^\ast = \argmin_\theta \langle \psi(\theta) | H | \psi(\theta)\rangle.
   \end{aligned}
\end{equation}
\par
A common way to achieve this parameterization is by introducing a parameterized unitary, $U(\theta)$, so that
    $|\psi\rangle = U(\theta)|0\rangle$
In the context of VQE, we call the parameterized circuit that implements $U(\theta)$ the \textit{ansatz}. Now, the ground state $E_0$ can be approximated by solving the following minimization problem 
\begin{equation}\label{exp}
\begin{aligned}
    E_0 \approx \min  \langle 0 |U^\dagger (\theta)HU(\theta)|0\rangle
\end{aligned}
\end{equation}
The particular choice of ansatz is of paramount importance to the performance of VQE \cite{Du_2022}.
\par
To actually find the optimal parameters $\theta^\ast$, VQE employs a classical optimizer. The expectation value in Equation \ref{exp} is computed on a quantum device, and this value alongside the current parameters are passed into a blackbox optimizer which then provides information to update $\theta$. This process is repeated iteratively until a desirable approximation of the ground state is obtained and the algorithm terminates. 
\subsection{CAFQA}
Because of the noise present in near term intermediate scale (NISQ) quantum devices, the performance of variational algorithms is often insufficient for practical problems~\cite{wang2021error, kandala2017hardware, ravi2021vaqem}. One technique for improving the noise robustness of VQAs is finding suitable initialization for the ansatz parameters. By doing so, the required number of optimization iterations can be reduced— ultimately reducing the number of times the noisy quantum circuit has to be executed. \par This approach has been previously studied by the CAFQA (Clifford Ansatz For Quantum Accuracy) protocol \cite{ravi2023cafqa, bhattacharyya2023optimal}. CAFQA is an initialization scheme that executes a classically simulatable search over a discrete parameter space to find appropriate initial values, and then initializes the continuous parameter optimization loop of a VQA with these  values.
\par 
In particular, CAFQA restricts its search space to parameter values for which the ansatz circuit is Clifford. It is well known that  Clifford circuits can be simulated in polynomial time \cite{Aaronson_2004}.
In practice, if a circuit consists of entirely Clifford gates (S, CNOT, H) and rotation gates (RX, RY, RZ), if all of the rotation angles are set to integer multiples of $\pi/2$, the corresponding circuit is Clifford.
\par 
Thus, for any ansatz where $\theta$ corresponds to the parameters of rotation gates, restricting the components of $\theta$ to integer multiples of $\pi/2$ effectively restricts the ansatz to Clifford circuits. Now, a discrete search can be executed in this restricted parameter space to find initial parameters for a full VQE run. 
\subsection{Variational Quantum Deflation}
Given a Hamiltonian $H$, an ansatz implementing $U(\theta)$ and an integer $k\geq 1$, Variational Quantum Deflation (VQD) is a VQA that aims to find the first $k$ eigenstates of the Hamiltonian $H$ \cite{Higgott_2019}.
\par
To do so, we first find the ground state $|\psi_0\rangle$ by solving a VQE problem. Then, we use $U(\theta^\ast_0)|0\rangle =|\psi_0\rangle$ to construct a new cost function
\begin{equation}
L_1(\theta) = \langle 0 | U^\dagger(\theta) H U(\theta)|0\rangle + \beta |\langle \psi_0 | U(\theta)|0\rangle|^2
\end{equation}
This cost function has two terms, the first is the standard VQE expectation value, but the second term constrains $U(\theta)|0\rangle$ to minimize its inner product with $|\psi_0\rangle$. The term $\beta$ is a hyperparameter which has to be chosen to be sufficiently large without dramatically increasing the difficulty of optimization. Because the eigenvalues of $H$ are mutually orthogonal, minimizing the parameters $\theta^*_1=\argmin L_1(\theta)$ satisfy $U(\theta^*_1)|0\rangle = |\psi_1\rangle$, providing the first excited state.
\par
This process can be repeated recursively, constructing 
\begin{equation}
    L_{n}(\theta)=\langle 0 | U^\dagger (\theta) H U(\theta)|0\rangle + \sum_{i=0}^{n-1} \beta_i |\langle \psi_i | U(\theta)|0\rangle|^2
\end{equation}
Which can then be optimized to find the $n$th excited state using $\theta^\ast_n = \argmin L_n(\theta)$ and $U(\theta^\ast_n)|0\rangle=|\psi_n\rangle$.
\\\\
\section{CAFQA VQD}
The most natural way to extend VQD to incorporate Clifford initialization is by restricting the states $|\psi_i\rangle$, to be a Clifford state.
\\\\
We define $|\phi_0\rangle$ to be the state that minimizes the expectation $\langle \phi | H | \phi\rangle$ over all Clifford states $| \phi\rangle$ obtainable $U(\theta)|0\rangle$. We then define the corresponding Clifford loss functions $$\tilde{L}_n(|\phi\rangle) = \langle \phi | H | \phi \rangle + \sum_i \beta_i |\langle \phi_i | \phi \rangle|^2$$
And define $|\phi_n\rangle$ to be the stabilizer state obtainable by $U(\theta)$ that minimizes the cost function $\tilde{L}_n$.
\\\\
All of these optimizations can be carried out using a classical device, because just like expectation values, the inner product of stabilizer states can be computed efficiently using their corresponding stabilizer groups \cite{garcia2013efficient}.
\begin{figure}[htbp]
    \centering
    \begin{subfigure}[b]{\linewidth}
        \includegraphics[width=\linewidth]{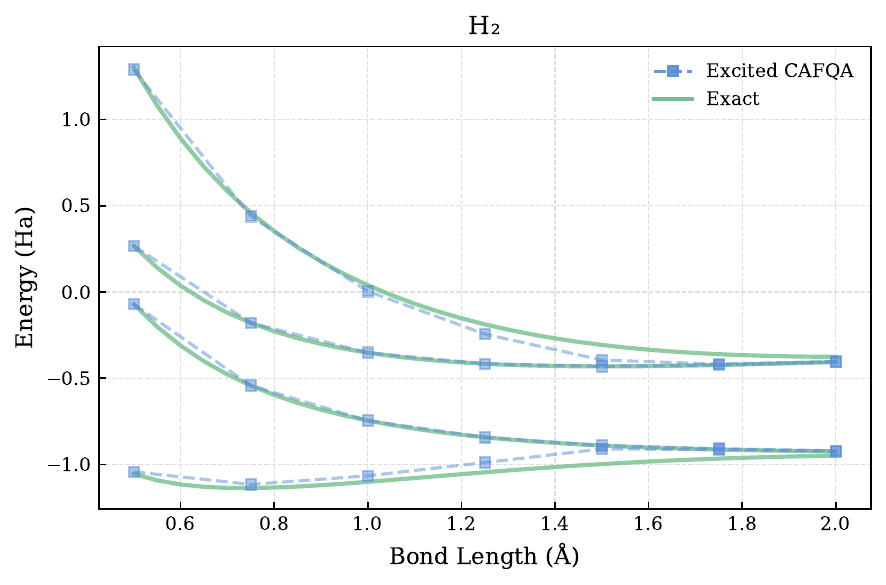}
        \caption{}
        \label{LeftMPS}
    \end{subfigure}%
    \hfill
    \begin{subfigure}[b]{\linewidth}
        \centering
        \includegraphics[width=\linewidth]{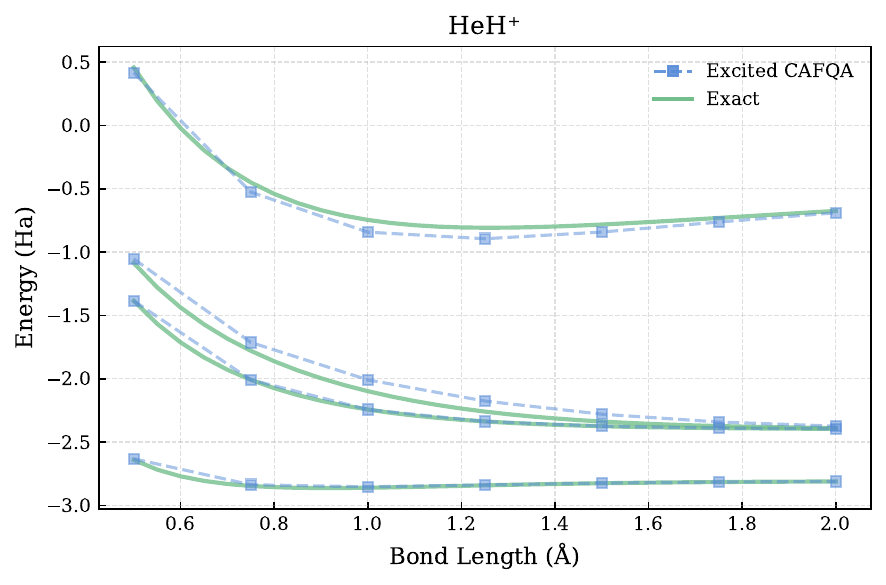}
        \caption{}
        \label{RightMPS}
    \end{subfigure}
   
    \caption{Comparison of exact and excited-CAFQA energies for (a) $\textrm{H}_2$ and (b) $\textrm{HeH}^+$ Hamiltonians. Code is available at \cite{Bhattacharyya_Excited_CAFQA}.}
    \label{plot}
\end{figure}
\par 
\begin{figure*}[htb]
    \centering
    \definecolor{rotationcolor}{rgb}{0.35, 0.55, 0.85} 
\definecolor{cnotcolor}{rgb}{0.45, 0.75, 0.55} 

\begin{quantikz}
    \gate[style={fill=rotationcolor!30}]{R_y(\theta_0)} & \gate[style={fill=rotationcolor!30}]{R_z(\theta_2)} & \ctrl[style={color=cnotcolor}]{1} &\gate[style={fill=rotationcolor!30}]{R_y(\theta_4)} & \gate[style={fill=rotationcolor!30}]{R_z(\theta_6)} & \ctrl[style={color=cnotcolor}]{1} &\gate[style={fill=rotationcolor!30}]{R_y(\theta_8)} & \gate[style={fill=rotationcolor!30}]{R_z(\theta_{10})}  &  \\
    \gate[style={fill=rotationcolor!30}]{R_y(\theta_1)} & \gate[style={fill=rotationcolor!30}]{R_z(\theta_3)} & \targ[style={color=cnotcolor}]{} 
    &\gate[style={fill=rotationcolor!30}]{R_y(\theta_5)} & \gate[style={fill=rotationcolor!30}]{R_z(\theta_7)} & \targ[style={color=cnotcolor}]{} 
    &\gate[style={fill=rotationcolor!30}]{R_y(\theta_9)} & \gate[style={fill=rotationcolor!30}]{R_z(\theta_{11})}  & 
\end{quantikz}
    \caption{Optimization ansatz}
    \label{ansatz}
\end{figure*}
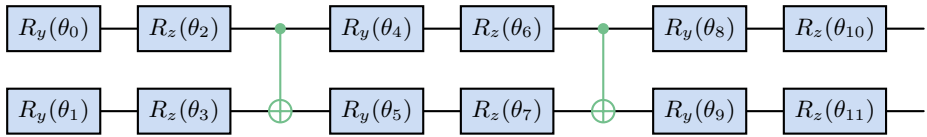
As a concrete example, we show the performance of CAFQA-VQD on $\textrm{H}_2$ and $\textrm{HeH}^+$ molecules, both of which correspond to $2$-qubit Hamiltonians after application of the Jordan Wigner transformation and a parity reduction. The results are shown in Fig. \ref{plot}. The maximum error for  the $\textrm{H}_2$ and $\textrm{HeH}^+$ problems were $0.087\textrm{ }\si{Ha}$ and $0.097\textrm{ }\si{Ha}$ respectively. Similar to the results from the original CAFQA optimization, the Clifford approximations for the energies converge to the exact at both small and large bond limits, while the maximum error occurs near the equilibrium (minimal energy) bond length. The optimizer used was essentially identical to that in \cite{ravi2023cafqa}, a black-box Bayesian optimizer. The only notable difference is that we use parameter transfer, that is, once we final optimal parameters for a given bond length, we use those parameters as an initial point for the next bond length \cite{Skogh2023Accelerating}. The circuit used for this optimization is shown in Fig. \ref{ansatz}.
\section*{Acknowledgement}
This material is based upon work supported by the U.S. Department of Energy, Office of Science, Office of Advanced Scientific Computing Research, Accelerated Research in Quantum Computing under Award Number DE-SC0025633.
\newpage

\bibliography{apssamp}

\begin{thebibliography}{12}%
\makeatletter
\providecommand \@ifxundefined [1]{%
 \@ifx{#1\undefined}
}%
\providecommand \@ifnum [1]{%
 \ifnum #1\expandafter \@firstoftwo
 \else \expandafter \@secondoftwo
 \fi
}%
\providecommand \@ifx [1]{%
 \ifx #1\expandafter \@firstoftwo
 \else \expandafter \@secondoftwo
 \fi
}%
\providecommand \natexlab [1]{#1}%
\providecommand \enquote  [1]{``#1''}%
\providecommand \bibnamefont  [1]{#1}%
\providecommand \bibfnamefont [1]{#1}%
\providecommand \citenamefont [1]{#1}%
\providecommand \href@noop [0]{\@secondoftwo}%
\providecommand \href [0]{\begingroup \@sanitize@url \@href}%
\providecommand \@href[1]{\@@startlink{#1}\@@href}%
\providecommand \@@href[1]{\endgroup#1\@@endlink}%
\providecommand \@sanitize@url [0]{\catcode `\\12\catcode `\$12\catcode `\&12\catcode `\#12\catcode `\^12\catcode `\_12\catcode `\%12\relax}%
\providecommand \@@startlink[1]{}%
\providecommand \@@endlink[0]{}%
\providecommand \url  [0]{\begingroup\@sanitize@url \@url }%
\providecommand \@url [1]{\endgroup\@href {#1}{\urlprefix }}%
\providecommand \urlprefix  [0]{URL }%
\providecommand \Eprint [0]{\href }%
\providecommand \doibase [0]{https://doi.org/}%
\providecommand \selectlanguage [0]{\@gobble}%
\providecommand \bibinfo  [0]{\@secondoftwo}%
\providecommand \bibfield  [0]{\@secondoftwo}%
\providecommand \translation [1]{[#1]}%
\providecommand \BibitemOpen [0]{}%
\providecommand \bibitemStop [0]{}%
\providecommand \bibitemNoStop [0]{.\EOS\space}%
\providecommand \EOS [0]{\spacefactor3000\relax}%
\providecommand \BibitemShut  [1]{\csname bibitem#1\endcsname}%
\let\auto@bib@innerbib\@empty
\bibitem [{\citenamefont {Tilly}\ \emph {et~al.}(2022)\citenamefont {Tilly}, \citenamefont {Chen}, \citenamefont {Cao}, \citenamefont {Picozzi}, \citenamefont {Setia}, \citenamefont {Li}, \citenamefont {Grant}, \citenamefont {Wossnig}, \citenamefont {Rungger}, \citenamefont {Booth},\ and\ \citenamefont {Tennyson}}]{Tilly_2022}%
  \BibitemOpen
  \bibfield  {author} {\bibinfo {author} {\bibfnamefont {J.}~\bibnamefont {Tilly}}, \bibinfo {author} {\bibfnamefont {H.}~\bibnamefont {Chen}}, \bibinfo {author} {\bibfnamefont {S.}~\bibnamefont {Cao}}, \bibinfo {author} {\bibfnamefont {D.}~\bibnamefont {Picozzi}}, \bibinfo {author} {\bibfnamefont {K.}~\bibnamefont {Setia}}, \bibinfo {author} {\bibfnamefont {Y.}~\bibnamefont {Li}}, \bibinfo {author} {\bibfnamefont {E.}~\bibnamefont {Grant}}, \bibinfo {author} {\bibfnamefont {L.}~\bibnamefont {Wossnig}}, \bibinfo {author} {\bibfnamefont {I.}~\bibnamefont {Rungger}}, \bibinfo {author} {\bibfnamefont {G.~H.}\ \bibnamefont {Booth}},\ and\ \bibinfo {author} {\bibfnamefont {J.}~\bibnamefont {Tennyson}},\ }\bibfield  {title} {\bibinfo {title} {The variational quantum eigensolver: A review of methods and best practices},\ }\href {https://doi.org/10.1016/j.physrep.2022.08.003} {\bibfield  {journal} {\bibinfo  {journal} {Physics Reports}\ }\textbf {\bibinfo {volume} {986}},\ \bibinfo {pages} {1–128} (\bibinfo {year}
  {2022})}\BibitemShut {NoStop}%
\bibitem [{\citenamefont {Du}\ \emph {et~al.}(2022)\citenamefont {Du}, \citenamefont {Tu}, \citenamefont {Yuan},\ and\ \citenamefont {Tao}}]{Du_2022}%
  \BibitemOpen
  \bibfield  {author} {\bibinfo {author} {\bibfnamefont {Y.}~\bibnamefont {Du}}, \bibinfo {author} {\bibfnamefont {Z.}~\bibnamefont {Tu}}, \bibinfo {author} {\bibfnamefont {X.}~\bibnamefont {Yuan}},\ and\ \bibinfo {author} {\bibfnamefont {D.}~\bibnamefont {Tao}},\ }\bibfield  {title} {\bibinfo {title} {Efficient measure for the expressivity of variational quantum algorithms},\ }\bibfield  {journal} {\bibinfo  {journal} {Physical Review Letters}\ }\textbf {\bibinfo {volume} {128}},\ \href {https://doi.org/10.1103/physrevlett.128.080506} {10.1103/physrevlett.128.080506} (\bibinfo {year} {2022})\BibitemShut {NoStop}%
\bibitem [{\citenamefont {Wang}\ \emph {et~al.}(2021)\citenamefont {Wang}, \citenamefont {Czarnik}, \citenamefont {Arrasmith}, \citenamefont {Cerezo}, \citenamefont {Cincio},\ and\ \citenamefont {Coles}}]{wang2021error}%
  \BibitemOpen
  \bibfield  {author} {\bibinfo {author} {\bibfnamefont {S.}~\bibnamefont {Wang}}, \bibinfo {author} {\bibfnamefont {P.}~\bibnamefont {Czarnik}}, \bibinfo {author} {\bibfnamefont {A.}~\bibnamefont {Arrasmith}}, \bibinfo {author} {\bibfnamefont {M.}~\bibnamefont {Cerezo}}, \bibinfo {author} {\bibfnamefont {L.}~\bibnamefont {Cincio}},\ and\ \bibinfo {author} {\bibfnamefont {P.~J.}\ \bibnamefont {Coles}},\ }\href@noop {} {\bibinfo {title} {Can error mitigation improve trainability of noisy variational quantum algorithms?}} (\bibinfo {year} {2021}),\ \Eprint {https://arxiv.org/abs/2109.01051} {arXiv:2109.01051 [quant-ph]} \BibitemShut {NoStop}%
\bibitem [{\citenamefont {Kandala}\ \emph {et~al.}(2017)\citenamefont {Kandala}, \citenamefont {Mezzacapo}, \citenamefont {Temme}, \citenamefont {Takita}, \citenamefont {Brink}, \citenamefont {Chow},\ and\ \citenamefont {Gambetta}}]{kandala2017hardware}%
  \BibitemOpen
  \bibfield  {author} {\bibinfo {author} {\bibfnamefont {A.}~\bibnamefont {Kandala}}, \bibinfo {author} {\bibfnamefont {A.}~\bibnamefont {Mezzacapo}}, \bibinfo {author} {\bibfnamefont {K.}~\bibnamefont {Temme}}, \bibinfo {author} {\bibfnamefont {M.}~\bibnamefont {Takita}}, \bibinfo {author} {\bibfnamefont {M.}~\bibnamefont {Brink}}, \bibinfo {author} {\bibfnamefont {J.~M.}\ \bibnamefont {Chow}},\ and\ \bibinfo {author} {\bibfnamefont {J.~M.}\ \bibnamefont {Gambetta}},\ }\bibfield  {title} {\bibinfo {title} {Hardware-efficient variational quantum eigensolver for small molecules and quantum magnets},\ }\href@noop {} {\bibfield  {journal} {\bibinfo  {journal} {Nature}\ }\textbf {\bibinfo {volume} {549}},\ \bibinfo {pages} {242} (\bibinfo {year} {2017})}\BibitemShut {NoStop}%
\bibitem [{\citenamefont {Ravi}\ \emph {et~al.}(2021)\citenamefont {Ravi}, \citenamefont {Smith}, \citenamefont {Gokhale}, \citenamefont {Mari}, \citenamefont {Earnest}, \citenamefont {Javadi-Abhari},\ and\ \citenamefont {Chong}}]{ravi2021vaqem}%
  \BibitemOpen
  \bibfield  {author} {\bibinfo {author} {\bibfnamefont {G.~S.}\ \bibnamefont {Ravi}}, \bibinfo {author} {\bibfnamefont {K.~N.}\ \bibnamefont {Smith}}, \bibinfo {author} {\bibfnamefont {P.}~\bibnamefont {Gokhale}}, \bibinfo {author} {\bibfnamefont {A.}~\bibnamefont {Mari}}, \bibinfo {author} {\bibfnamefont {N.}~\bibnamefont {Earnest}}, \bibinfo {author} {\bibfnamefont {A.}~\bibnamefont {Javadi-Abhari}},\ and\ \bibinfo {author} {\bibfnamefont {F.~T.}\ \bibnamefont {Chong}},\ }\href@noop {} {\bibinfo {title} {Vaqem: A variational approach to quantum error mitigation}} (\bibinfo {year} {2021}),\ \Eprint {https://arxiv.org/abs/2112.05821} {arXiv:2112.05821 [quant-ph]} \BibitemShut {NoStop}%
\bibitem [{\citenamefont {Ravi}\ \emph {et~al.}(2023)\citenamefont {Ravi}, \citenamefont {Gokhale}, \citenamefont {Ding}, \citenamefont {Kirby}, \citenamefont {Smith}, \citenamefont {Baker}, \citenamefont {Love}, \citenamefont {Hoffmann}, \citenamefont {Brown},\ and\ \citenamefont {Chong}}]{ravi2023cafqa}%
  \BibitemOpen
  \bibfield  {author} {\bibinfo {author} {\bibfnamefont {G.~S.}\ \bibnamefont {Ravi}}, \bibinfo {author} {\bibfnamefont {P.}~\bibnamefont {Gokhale}}, \bibinfo {author} {\bibfnamefont {Y.}~\bibnamefont {Ding}}, \bibinfo {author} {\bibfnamefont {W.~M.}\ \bibnamefont {Kirby}}, \bibinfo {author} {\bibfnamefont {K.~N.}\ \bibnamefont {Smith}}, \bibinfo {author} {\bibfnamefont {J.~M.}\ \bibnamefont {Baker}}, \bibinfo {author} {\bibfnamefont {P.~J.}\ \bibnamefont {Love}}, \bibinfo {author} {\bibfnamefont {H.}~\bibnamefont {Hoffmann}}, \bibinfo {author} {\bibfnamefont {K.~R.}\ \bibnamefont {Brown}},\ and\ \bibinfo {author} {\bibfnamefont {F.~T.}\ \bibnamefont {Chong}},\ }\href@noop {} {\bibinfo {title} {Cafqa: A classical simulation bootstrap for variational quantum algorithms}} (\bibinfo {year} {2023}),\ \Eprint {https://arxiv.org/abs/2202.12924} {arXiv:2202.12924 [quant-ph]} \BibitemShut {NoStop}%
\bibitem [{\citenamefont {Bhattacharyya}\ and\ \citenamefont {Ravi}(2023)}]{bhattacharyya2023optimal}%
  \BibitemOpen
  \bibfield  {author} {\bibinfo {author} {\bibfnamefont {B.}~\bibnamefont {Bhattacharyya}}\ and\ \bibinfo {author} {\bibfnamefont {G.~S.}\ \bibnamefont {Ravi}},\ }\bibfield  {title} {\bibinfo {title} {Optimal clifford initial states for ising hamiltonians},\ }in\ \href@noop {} {\emph {\bibinfo {booktitle} {2023 IEEE International Conference on Rebooting Computing (ICRC)}}}\ (\bibinfo {organization} {IEEE},\ \bibinfo {year} {2023})\ pp.\ \bibinfo {pages} {1--10}\BibitemShut {NoStop}%
\bibitem [{\citenamefont {Aaronson}\ and\ \citenamefont {Gottesman}(2004)}]{Aaronson_2004}%
  \BibitemOpen
  \bibfield  {author} {\bibinfo {author} {\bibfnamefont {S.}~\bibnamefont {Aaronson}}\ and\ \bibinfo {author} {\bibfnamefont {D.}~\bibnamefont {Gottesman}},\ }\bibfield  {title} {\bibinfo {title} {Improved simulation of stabilizer circuits},\ }\bibfield  {journal} {\bibinfo  {journal} {Physical Review A}\ }\textbf {\bibinfo {volume} {70}},\ \href {https://doi.org/10.1103/physreva.70.052328} {10.1103/physreva.70.052328} (\bibinfo {year} {2004})\BibitemShut {NoStop}%
\bibitem [{\citenamefont {Higgott}\ \emph {et~al.}(2019)\citenamefont {Higgott}, \citenamefont {Wang},\ and\ \citenamefont {Brierley}}]{Higgott_2019}%
  \BibitemOpen
  \bibfield  {author} {\bibinfo {author} {\bibfnamefont {O.}~\bibnamefont {Higgott}}, \bibinfo {author} {\bibfnamefont {D.}~\bibnamefont {Wang}},\ and\ \bibinfo {author} {\bibfnamefont {S.}~\bibnamefont {Brierley}},\ }\bibfield  {title} {\bibinfo {title} {Variational quantum computation of excited states},\ }\href {https://doi.org/10.22331/q-2019-07-01-156} {\bibfield  {journal} {\bibinfo  {journal} {Quantum}\ }\textbf {\bibinfo {volume} {3}},\ \bibinfo {pages} {156} (\bibinfo {year} {2019})}\BibitemShut {NoStop}%
\bibitem [{\citenamefont {Garcia}\ \emph {et~al.}(2013)\citenamefont {Garcia}, \citenamefont {Markov},\ and\ \citenamefont {Cross}}]{garcia2013efficient}%
  \BibitemOpen
  \bibfield  {author} {\bibinfo {author} {\bibfnamefont {H.~J.}\ \bibnamefont {Garcia}}, \bibinfo {author} {\bibfnamefont {I.~L.}\ \bibnamefont {Markov}},\ and\ \bibinfo {author} {\bibfnamefont {A.~W.}\ \bibnamefont {Cross}},\ }\href@noop {} {\bibinfo {title} {Efficient inner-product algorithm for stabilizer states}} (\bibinfo {year} {2013}),\ \Eprint {https://arxiv.org/abs/1210.6646} {arXiv:1210.6646 [cs.ET]} \BibitemShut {NoStop}%
\bibitem [{\citenamefont {Bhattacharyya}()}]{Bhattacharyya_Excited_CAFQA}%
  \BibitemOpen
  \bibfield  {author} {\bibinfo {author} {\bibfnamefont {B.}~\bibnamefont {Bhattacharyya}},\ }\href {https://github.com/BBhattacharyya1729/ExcitedCAFQA-} {\bibinfo {title} {{Excited CAFQA}}}\BibitemShut {NoStop}%
\bibitem [{\citenamefont {Skogh}\ and\ \citenamefont {Rahm}(2023)}]{Skogh2023Accelerating}%
  \BibitemOpen
  \bibfield  {author} {\bibinfo {author} {\bibfnamefont {M.}~\bibnamefont {Skogh}}\ and\ \bibinfo {author} {\bibfnamefont {M.}~\bibnamefont {Rahm}},\ }\href {https://doi.org/10.5878/fvza-z272} {\bibinfo {title} {Accelerating variational quantum eigensolver convergence using parameter transfer}} (\bibinfo {year} {2023})\BibitemShut {NoStop}%
\end{thebibliography}%

\end{document}